# High speed and reconfigurable optronic neural network with digital nonlinear activation

Qiuhao Wu[1], Jia Liu[1], Xiubao Sui[1]*, Qian Chen[1], and Liping Wang[1]

[1]*School of Electronic and Optical Engineering, Nanjing University of Science and Technology, Nanjing 210094, China*

E-mail: sxbhandsome@163.com

With its unique parallel processing capability, optical neural network has shown low-power consumption in image recognition and speech processing. At present, the manufacturing technology of programmable photonic chip is not mature, and the realization of optical neural network in free-space is still a hot spot of intelligent optical computing. In this article, based on MNIST datasets and 4f system, three-layer optical neural networks are constructed, whose recognition accuracy can reach 93.66%. Our network is programmable, high speed, reconfigurable and is better than the existing free-space optical neural network in terms of spatial complexity.

In the era of big data, more requirements are put forward for the information processing capacity of computers, which gives rise to the research of quantum computing,[1] optical computing,[2] and other emerging fields of supercomputing.[3-4] After half a century of development, artificial neural network (ANN) has shown excellent performance in image processing,[5-6] speech recognition,[7] three-dimensional imaging,[8-9] target detection,[10-12] etc. However, the demand of computing power is becoming more and more stringent. Therefore, the mode of signal transmission in neural network is transformed from electrical to optical. It is well known that using light as the signal transmission medium enables the system to have the ability of large-scale parallel processing and high bandwidth.[13-15] In recent years, the research of optical neural network (ONN) has become more and more popular. At present, it has achieved success in natural language processing,[16] handwriting digit recognition,[17-23] pedestrian detection,[24] intelligent sensor design and other fields.[25] However, their architectures are complex and the recognition accuracy needs to be improved. As one of the most widely used architectures in computer vision, convolutional neural network (CNN) utilizes filters to extract image features.[26] Here, we use optical signal convolution theory,[27] based on the digital nonlinear activation in the free-space, optronic neural network is constructed, and the accuracy of identification can reach 93.66%. Compared to the convolutional opto-electrical hybrid neural network relied on digital neural network,[28-29] our architecture makes full use of the parallel processing ability of the light, the whole system has lower spatial complexity and is reconfigurable.

As shown in Fig.1(a), the neural network realizes its function through the cross connection of neurons and by means of the back-propagation algorithm.[30] The core of its work includes matrix multiplication, activation and objective function. The process of constructing ONN can be regarded as finding suitable optical devices to transplant the function of digital neural network into optical system. It is known from physical optics that light has wave-particle duality. Based on the superposition principle of diffractive wavelet,[31] X. Lin *et al.* analogized the cross interconnection of neurons to realize information transmission, and thus constructed an all-optical diffractive neural network,[18] as shown in Fig.1(b). Different from the superposition principle of wavelet, we use 4f system to realize matrix multiplication, as we mentioned in the review of ONN.[32-33] In addition, the diffractive network makes the parameters of trained model into the diffractive layer, which makes the function of the network fixed and lacks reconfigurability. In this paper, the optronic neural network proposed by us is programmable and reconfigurable due to the use of high-speed digital mirror device (DMD) and high-speed camera.

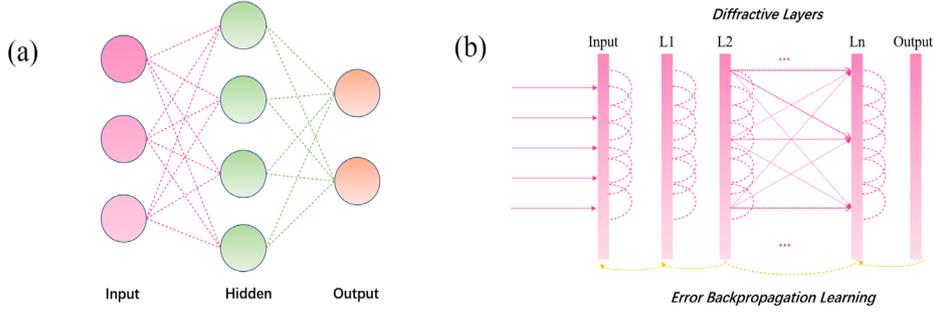

**Fig.1.** (a) The structure of nanophotonic neural network: $X_{n+1} = F(X_n \cdot (W_{n+1} + B_n))$, F is the nonlinear activation function. (b) All-optical Diffractive Neural Network: $X_{n+1} = W_{n+1} \cdot (X_n \circ B_n)$.

Inspired by the 1×1 full CNN and to improve the accuracy of the model, we initialize a weight matrix W of the same size as the input image (512×512), multiply the corresponding elements of the input, and add a bias matrix B to obtain the pre-processed image. This can be realized by MZI device in optical system, as shown in Fig.2(b). We know that for the superposition of two vector waves E₁ and E₂ which have an angle α between the direction of vibration. It can be described as

$$E(r,t) = E_1(r,t) + E_2(r,t) \tag{1}$$

The intensity of a vector field is the time average of its conjugate dot product

$$I = <E \cdot E^*> = <[E_1(r,t) + E_2(r,t)] \cdot [E_1^*(r,t) + E_2^*(r,t)]> \tag{2}$$

It can be further written as

$$\begin{aligned} I &= E_1 \cdot E_1^* + E_2 \cdot E_2^* + <\mathrm{Re}\{2E_1 \cdot E_2^*\}> \\ &= A_1^2 + A_2^2 + 2A_1 A_2 \cos\alpha <\cos\psi> \\ &= I_1 + I_2 + I_{12} \end{aligned} \tag{3}$$

It can be seen from formula (3) that when α=90°, namely, when the vibration directions of two waves are perpendicular to each other, the coherent term $I_{12}$ is zero, and the intensity effect of superposition is shown as the summation of their respective intensities. This is just enough to optically realized WX+B, what we called preprocessing.

The preprocessed image is fed into the DMD and is optical convolved with the initial phase mask on the Fourier plane by the optical 4f system. The output light field after Fourier inversion is received by the detector placed on the focal plane of the lens. This enables the forward propagation of optical signal in our proposed architecture. As shown in Fig.2(a). The output of sCMOS is then activated by the digital nonlinear activation such as ReLu, SoftMax. In addition, the number of layers in our network is scalable. Researchers can then feed the output of Computer into the DMD to increase the number of layers in the network and need to imprint the phase parameters of the new layer into the spatial light modulator (SLM).

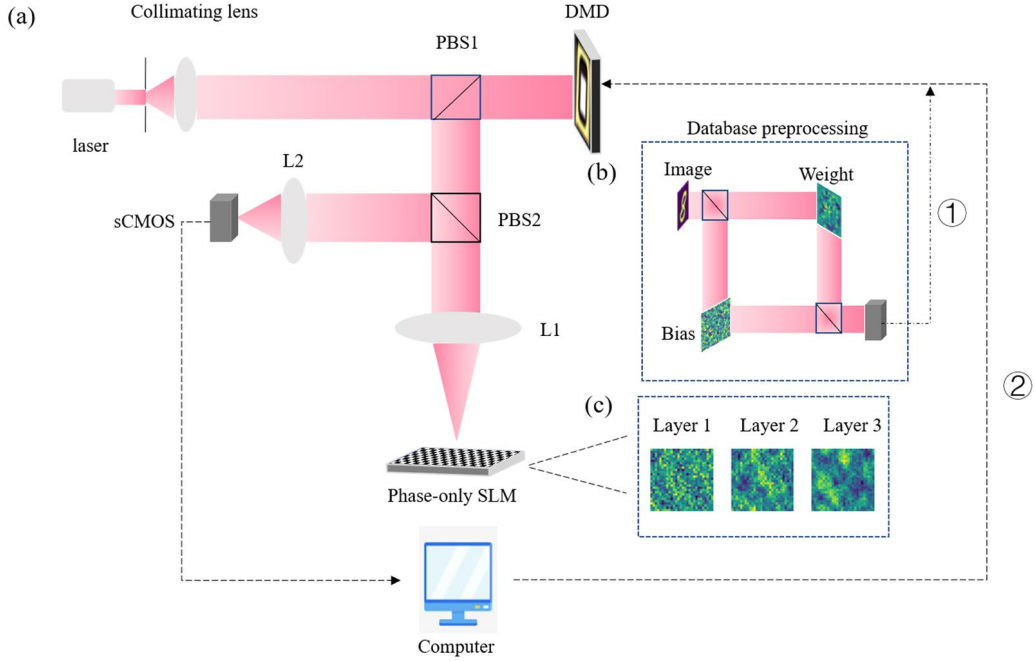

**Fig.2.** The schematic diagram of three-layers ONN. (a) The pre-processed input image is received by the detector through the optical 4f system, and the process can be expressed as $O = \mathcal{F}^{-1}\{\mathcal{F}(X \circ W + B) \circ \phi)\}$. Here, $\mathcal{F}$ means Fourier transform, ∘ means elements multiplication. O: the output field of the last layer. In addition, the symbol * in figure equals to ∘ in equation and the symbol + can realized by MZI based on superposition of non-interference. (b) Preprocessing: use MZI to realize WX+B. (c) The phase mask of each layer which is imprinted on the phase-only spatial light modulator.

Our goal is to identify the MNIST datasets to be tested, which contains 10 categories, so we divided 10 detected areas on the detector. If the input image is a handwritten digit 8 and the light intensity of the eighth detection area is the largest compared to other areas, it is considered that our architecture correctly recognizes the image to be tested. We know that any iterative algorithm can be trained and converged by a neural network model. In our algorithm, initialized weights W, bias B, and phase masks are all learnable variables. Finally, the ideal model was obtained by using 55,000 training sets, 5000 validation sets, and the generalization ability of the model was verified by 10,000 test sets. The algorithm includes forward propagation and back propagation, adopts stochastic gradient descent, cross entropy loss function, and $\alpha$ represents learning rate.

$$Pre = W \circ X + B \quad (4)$$

$$O = \mathcal{F}^{-1}\{\mathcal{F}(Pre) \circ \phi\} \quad (5)$$

$$I = O \cdot O^* \quad (6)$$

$$LOSS = -\sum_{i=1}^{n} \hat{I} \log(I) \quad (7)$$

$$W = W - \alpha \frac{\partial LOSS}{\partial W} \quad (8)$$

$$B = B - \alpha \frac{\partial LOSS}{\partial B} \quad (9)$$

$$\phi = \phi - \alpha \frac{\partial LOSS}{\partial \phi} \quad (10)$$

Based on the above algorithm, we set up the optronic neural network with one-layer and three-layer, respectively. The so-called 3-layer ONN is achieved by feeding the output of the Computer to the DMD and imprinting the new phase mask on the phase-only SLM, as shown in Fig.2(a). As we discussed in ref.31 and ref.32, nonlinear activation plays a key role in the multi-layer ONN. Here, we use the digital ReLu to realize the nonlinear activation. Considering the light intensity is non-negativity, we choose to shift the turning point of the function to fit it. In addition, the phase mask of each layer in our ONN is shown in Fig.2(c).

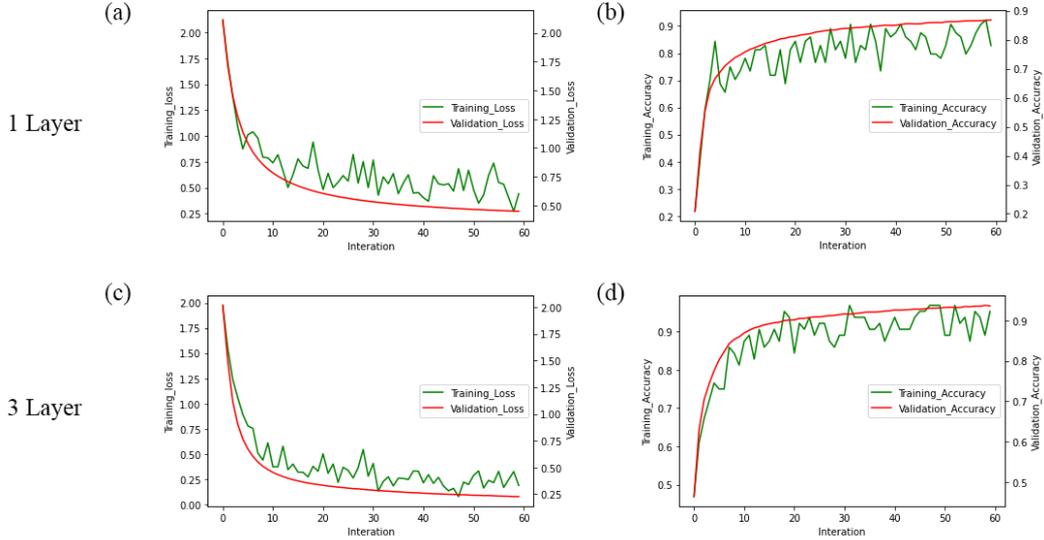

**Fig.3.** The results of the training for the one- and three-layer ONN with the size of 55000 training datasets and 5000 validation datasets. (a), (c) describe the training loss and validation loss of the ONN. (b), (d) show the training accuracy and validation accuracy of the ONN.

In view of the fact that we use optical 4f system to realize matrix multiplication in free space and digital nonlinear activation, we called this architecture as the free-space optronic neural network. Different from another research,[23] the algorithm is a crucial step for input image preprocessing by learning the optimal weight W and bias B, and ultimately reduces the number of layers of the network and optimizes the spatial complexity of the model. Similar to the previous studies on ONN, we also used MNIST database to design the optical neural network system. The performance of training is shown in Fig.3. In our architecture, the recognition accuracy of one- and three-layers reached 86.06% and 93.66%, respectively. In addition, Fig.4 shows the recognition accuracy of MNIST data.

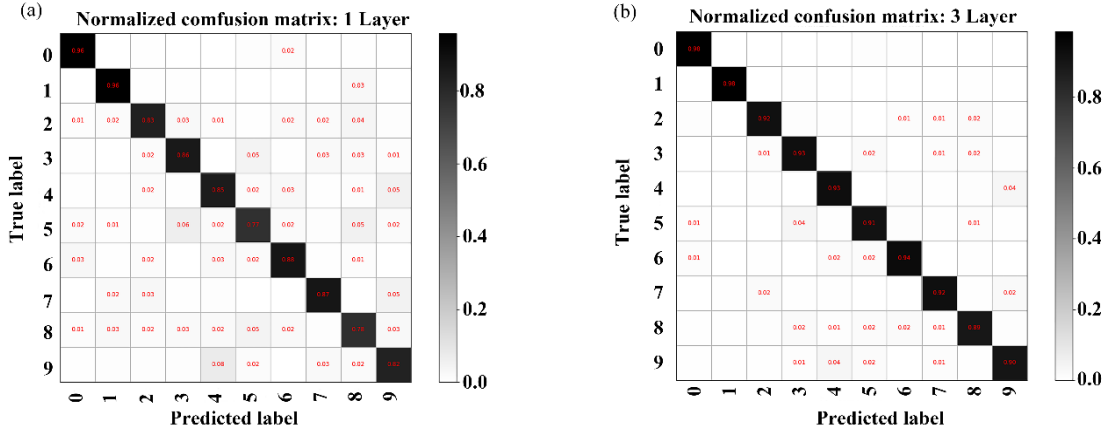

**Fig.4.** (a)-(b) show the test performance of the 10000 MNIST datasets, and the accuracy for one- and three-layer ONN is 86.06% and 93.66%, respectively.

It can be seen from Table I. that our 3-layer ONN has the acceptable recognition accuracy. We have a reason to believe that by increasing the number of layers of the network and adopting Regularization, Batch Norm, Shortcut connection, our architecture will have higher recognition accuracy.

Table I.   The test accuracy of MNIST in Optical Neural Networks.

| Research | Accuracy |
| --- | --- |
| *Ref.17* (Nano ONN) | 79% |
| *Ref.18* (5-layers D$^2$NN) | 91.75% |
| *Ref.18* (7-layers D$^2$NN) | 93.39% |
| *Ref.19* (5-layers ONN) | 88.2% |
| *Ref.21* (10-layers ONN) | 91.96% |
| *Ref.23* (6-layers FONN) | 92.51% |
| *Ours* (1-layer CONN) | 86.06% |
| *Ours* (3-layer CONN) | 93.66% |

We know the concept of using phase-masks in 4f system is well known and has also extensively been studied in the scope of neural networks, such as Ref.29 and Ref.30. In our opinions, the main method of Chang et al. is based on CNN, they use multi-kernal to extract features of the object and tail them in a 2D plane. Although they use an optical correlator similar to our 4f system, but their one optical correlator must be processed by the digital CNN or full-connected NN and then they obtain a good performance in MNIST (their accuracy lower than ours). However, the biggest difference of our architecture is that we use polarization to preprocess the input image, utilize computer to implement nonlinear activation and then pass through three spatial optical convolutional layers to obtain excellent recognition accuracy. On the one hand, our ONN is scalable and fast, the setups in Fig.1(a) can classify more complex datasets by adding additional layers. Considering the speed of light and the use of DMD and sCMOS, the latency of overall system is very low. On the other hand, the DMD and SLM are programmable, this enables the system has reconfigurability.

The number of neurons in each layer is 512×512, it can be seen as the nodes of the one-layer ONN. Total nodes of our network are 3×512×512, ~786000 nodes. In addition, a lens is a passive device, Fourier transform of which takes about 1ps to perform. If the parameters of our ONN have been learned, we then use the DMD and phase-only SLM to load the weight, bias and phase, whose high response rate (10 kHz) also ensures that our optical testing system responds very quickly. The latency of our single-layer convolution is only a few milliseconds. Considering our ONN have 3 layers and the $10^7$ HZ clock rate, which correspond to $7.86\times10^{12}$ dimensional multiplications per second. The number of operations (FLOPs) can be given by

$$N = 2\times 786000^2 \times 10^7 \, FLOPs \qquad (11)$$

The power consumption of our ONN is dominated by the laser, DMD, SLM and the digital computer, whose total power needed is approximately 25W. Therefore, the energy per FLOP of the CONN is $2\times 4\times 786000^2 \times 10^5 \, FLOPs/J$ ~4.94×$10^{17}$FLOPs/J.[16] As a result, our ONN has higher speed, much lower power consumption.

In summary, we implement an optronic neural network in free space with only three layers. We use DMD and phase-only SLM to load the learned parameters obtained from the neural network. Different from making diffractive layers or metasurfaces by fixed parameters, our system is programmable and scalable. However, there is still a gap between the present recognition accuracy and the expression ability of digital neural network. Future research will focus on the construction of all-optical neural network based on low power all-optical nonlinear activation function. At the same time, the adaptive online training algorithm is also need to be studied to enable the system to have the ability of real-time detection and recognition.

**Acknowledgments**

This work was supported in part by the National Natural Science Foundation of China (Grant Nos. 11773018 and 61727802), and in part by the Fundamental Research Funds for the Central Universities (Grant Nos. 30919011401 and 30920010001).